\documentclass[12pt]{article}
\usepackage{amsmath,amssymb}
\usepackage{graphicx,color}
\numberwithin{equation}{section}
\usepackage{cite}
\usepackage{bm}
\usepackage{dcolumn}
\newcommand{\be}{\begin{equation}}
\newcommand{\bea}{\begin{eqnarray}}
\newcommand{\eea}{\end{eqnarray}}
\newcommand{\ba}{\begin{array}}
\newcommand{\ea}{\end{array}}
\newcommand{\ee}{\end{equation}}

\expandafter\ifx\csname mathbbm\endcsname\relax

\else

\fi \textheight 22cm \textwidth 15cm \topmargin 1mm \oddsidemargin
5mm \evensidemargin 5mm

\begin{document}
\begin{titlepage}
\hfill \vbox{
    \halign{#\hfil         \cr
         \cr
                      } 
      }  
\vspace*{20mm}
\begin{center}
{\Large {\bf The Scalar Field Effective Action for The Spontaneous Symmetry Breaking in Gravity }\\
}

\vspace*{15mm} \vspace*{1mm} {Amin Akhavan}

\vspace*{.4cm}

{\it  School of Particles and Accelerators, Institute for Research in Fundamental Sciences (IPM)\\
P.O. Box 19395-5531, Tehran, Iran\\$email:amin_- akhavan@ipm.ir$ }

\vspace*{2cm}

\end{center}

\begin{abstract}
We calculate the quantum effective action for a scalar field which
has been recently used for a specific kind of symmetry breaking in
gravity. Our study consists of calculating the 1-loop  path integral
of canonical momentum and determining the renormalization
conditions. We will also discuss on the new renormalization
conditions to redefine the new degrees of freedom corresponding to a
massive vector field.

\end{abstract}

\vspace{2cm}

\end{titlepage}

\section{Introduction}
Studying the creation and formation of the universe is an important
area in gravity where the spontaneous symmetry breaking can have
useful application in\cite{englert}. Indeed, production of the
particles without loss of energy and arising the gauge bosons can be
regarded as the results of the symmetry breaking method. There is
also a sample of symmetry breaking which gives mass to the graviton.
A massive graviton leads the newtonian gravity force to fall down at
large distances. In addition, studying of the massive gravity and
the mutual effects between the mass and the cosmological constant
and also investigating the stability and renormalizability problems
are another points of interest in the massive gravity framework.

A covariant higgs mechanism method in gravity and its unitatrity
problem have also been explained by Gerard t'Hooft\cite{Gerard}. In
the t'Hooft formalism, four scalar fields $\phi^A(x), A=1,...,4$, have
been used to break the gauge symmetry. These fields which have the
global symmetry $so(3,1)$, enter in the action through the usual
quadratic form: \be
\sqrt{-g}g^{\mu\nu}\partial_{\mu}\phi^A(x)\partial_{\nu}\phi^A(x)\nonumber
\ee

In the presence of the cosmological constant, we can choose the
solution in the form of $\phi^A(x)\propto m\delta^A_{\mu}x^{\mu}$
(as a dynamical vacuum) to produce the mass terms for all gravity
degrees of freedom. In this way the symmetry of diffeomorphism
reduces to the global $so(3,1)$. One of the massive degrees of
freedom has a kinetic term with different sign in the Ricci scalar.
Such a difference makes the classical solution unstable and also the
unitarity failed.

There has been made some attempts to study the massive gravity
solution which none of them needed to solve the unitatrity
problem\cite{veltman,fierz}. However in the t'Hooft version of
massive gravity, the unitatrity problem has to be solved. In the
t'Hooft method, there are six massive degrees of freedom: Five
traceless $\overline{h}_{ij}(x)$ with definite kinetic terms. They
play the role of degrees of freedom for a spin 2 particle. The
remaining degree of freedom is a scalar $u(x)$ corresponding to a
scalar particle. As have pointed out above, its kinetic term is
indefinite and the solution is unstable. In order to restore the
untarity, $u(x)$ has to be decoupled from the other matter fields.

Although the mass of $u(x)$ in the Fierz-Pauli approach is
infinite\cite{fierz}, it is  finite in the t'Hooft mechanism and
therefore has to be decoupled through a different way. The proposal
is to define a coupling between the matter fields and a new metric
$g_{\mu\nu}^{matter}$ which is constructed from the original metric
and the Higgs scalar field. Such a coupling is defined so that
$u(x)$ doesn't couple to any other matter field. By using this
proposal, $u(x)$ is finally decoupled from the matter fields only in
the linear terms of $g_{\mu\nu}^{matter}$.

By use of the original metric and scalar fields, Chamseddin and
Mukhanov, defined a new metric as the following\cite{chams}, \be
H^{AB}(x)=g^{\mu\nu}(x)\partial_{\mu}\phi^A(x)\partial_{\nu}\phi^B(x)\nonumber
\ee They added a new term functional of $H^{AB}$ to the Einstein and
Hilbert action such that creates the Fierz and pauli mass terms.
Their higgs mechanism is clean of unitarity problem and also is
linearly ghost free. The action for the scalar fields that they
considered is: \be
S(\phi)\propto\int(d^4x\sqrt{-g})\,\,\,3(\frac{1}{16}H^2-1)^2
-\overline{H}^A_{B}\overline{H}^B_{A}\nonumber \ee Such that,
$H=H^A_A$ and $\overline{H}^A_B=H^A_B-\frac{\delta^A_B}{4}H$. In
such an action, since there exist objects like $\dot{\phi}^n$, $n>2$
, we require the new canonical process to have hamiltonian and
poisson brackets\cite{Kluson} and also to quantize the theory.

R. Fukuda and E. Kyriakopoulos, derived the effective potential
through the path integral with a constraint on the zero mode
field\cite{FUKUDA}. The effective potential is function of the zero
mode field, and the extremum of the effective potential is the
classical solution of the action. By using their method, the
symmetry breaking could be understood more
clearly\cite{CONSTRAINT,Wetterich}.

In this paper in section 2, we will use Fukuda and Kyriakopoulos
method to define the spontaneous symmetry breaking in the t'Hooft
mechanism. In section 3, the one-loop effective action will be
obtained from the actions with different functionality of canonical
momentum, like Chamseddine and Mukhanov sample. In section 4,
firstly we will consider the normal renormalization conditions to
practically calculate the effective action. Secondly In subsection
4.1, through the new renormalization conditions we will construct
the New degrees of freedom. Such degrees of freedom which represent
a non symmetric vector field in the IR scale.

\section{Spontaneous symmetry breaking}

In the quantum field theory, the spontaneous symmetry breaking puts
the quantum modes on a solution of the classical equation of motion.
In fact we exchange the quantum vacuum with a macroscopic state in
which the classical field could be measured in. A quantum vacuum is
a linear combination of all eigne states of the field. By expelling
the classical solution from the linear combination, we can define
the effective vacuum. By this definition, the symmetry in the
effective vacuum state will be missed. For example, considering a
lagrangian with even function of scalar field, we have: \be
\langle\Omega\mid\phi\mid\Omega\rangle\,\propto \int
\mathcal{D}\phi\,\phi \, e^{iS(\phi^2)}=0 \ee While by expelling the
zero momentum mode from the path integral, the symmetry
$\phi\rightarrow -\phi$ will be broken. \be
\langle\Omega^\prime\mid\phi\mid\Omega^\prime\rangle\propto \int
\mathcal{D}\phi\,\phi\, e^{iS(\phi^2)}\delta(\frac{1}{V}\int
dx\phi(x)-\phi_{0}) \neq 0 \ee As a
result,$\langle\Omega^\prime\mid\phi\mid\Omega^\prime\rangle$ is the
classical solution of the equation of motion of the scalar field,
which could be measured in the macroscopic scale. In fact such a
macroscopic scale also determines the renormalization conditions.
These conditions define the observable effective objects from the
fundamental unscreened objects. Through the averaging of all
macroscopic states, the expectation value of the field in the
original vacuum state will be obtained: \be
\langle\Omega\mid\phi\mid\Omega\rangle\,\propto\int
d\phi_{0}\,\langle\Omega^\prime\mid\phi\mid\Omega^\prime\rangle=0
\ee For the diffeomorphism symmetry breaking in gravity, the scalar
field lagrangian would be defined as a function of $H^{AB}$: \be
H^{AB}=g^{\mu\nu}\partial_{\mu}\phi^{A}\partial_{\nu}\phi^{B}
\nonumber \ee By choosing the classical solution $\phi^{A}\propto
x^{A}$  as an effective vacuum, the diffeomorphism will be broken
and the gravity will be massive. The Riemannian space $x^{\mu}$
becomes to a flat one $x^{A}$, and
the symmetry of diffeomorphism reduces to the lorentz symmetry.\\

To consider the separation between the quantum modes and the
classical solution, the partition function will be written like
this:

\bea
\mathcal{Z}=
\int da^{A}db_{\mu}^{B}\int \mathcal{D}\phi \mathcal{D}\pi
\delta(\frac{1}{V}\int dx \phi^{A}-a^{A})\delta(\frac{1}{V}\int
dx\partial_{\mu}\phi^{B}-b_{\mu}^{B})\nonumber\\
 \exp({i\int dx(\dot{\phi}\pi-\mathcal{H}(\phi,\pi))})
\eea By considering
$\phi^{A}(x)=a^{A}+b_{\mu}^{A}x^{\mu}+\chi^{A}(x)$, we can define
$\chi^A(x)$ as a fourier expansion. The fourier coefficients are
analytical functions of momentum like $\chi^{A}(p=0)=0$ ,
$\chi^{A}(p\neq0)<\infty$.
By defining a Hilbert space constructed of the states that $a$ and $b$ are measured in, a usual quantum theory will be created for the fields $\chi^{A}$.\\

Of course, if the lagrangian has only the gravitational probes, the
measurable classical values in the macroscopic scale are
$\partial_{\mu}\phi^{A}$ which are coupled to the gravitational
metric. And in such a situation, $\phi(x)$ is not measurable. But if
the zero mode of $\phi(x)$ to be determined in an experiment, it is
a result of a little term (functional of $\phi(x)$) that is attached
to the real lagrangian.
Such a little term, is omitted at the first order of approximation in the macroscopic scale. \\

Anyhow in the measured states with values $b_{\mu}^{A}$, we can use
a generating function for obtaining the quantum quantities: \be
e^{iF(J)}=\int \mathcal{D}\phi \mathcal{D}\pi\,
\delta(\frac{1}{V}\int dx\partial_{\mu}\phi^{A}-b_{\mu}^{A})
\exp({i\int dx(\dot{\phi}\pi-\mathcal{H}+iJ\phi)}) \ee

\section{Effective action}

In all lagrangians that are quadratic functions of time derivative
of fields, it can be shown that: \be \int \mathcal{D}\pi e^{i\int
dx(\dot{\phi}\pi-\mathcal{H}(\phi,\pi))}=e^{iS(\phi,\dot{\phi})} \ee
While the lagrangians are used to break the gauge symmetry in the
gravity are mainly functions of the higher power of time derivative,
therefore equation (3.1) loses its efficiency. In such lagrangians
to use the path integral method, The hamiltonian and the canonical
momentum have to be obtained and the integration over the canonical
momentum has to be calculated. In the general form, we can calculate
the path integral via the perturbation method around the extremum
point of the term inside the exponent: \be
\dot{\phi}(\phi(x),\pi(x))-\frac{\delta}{\delta \pi_{A}(x)}\int dx
\mathcal{H}(\phi,\pi)=0 \ee
That the extremum point of $\pi_{A}(x)$ is arrived from the time derivative of the scalar fields. \\
Through expansion of $\pi_{A}(x)$ around the extremum point, we can
obtain the generating function at the one-loop order: \bea
e^{iF(J)}= \int\mathcal{D}\phi \, e^{S(\phi,\dot{\phi})+i\int dx J(x)\phi(x)}\int \mathcal{D}\pi\nonumber\\
 \exp{-i\int dx dy(\pi_{A}(x)-\pi_{A}(\phi,\dot{\phi}))\frac{\delta^2 \int dz\mathcal{H}(\phi,\pi)}{\delta\pi_{A}(x)\delta\pi_{B}(x)}\mid_{\pi(\phi,\dot{\phi})}(\pi_{B}(x)-\pi_{B}(\phi,\dot{\phi}))}
\eea It has to be attended to the first integral that has been
written only in the space of the fourier modes of $\phi(x)$. In the
equation (3.3), the term $Lndet(\frac{\delta^2\int
dz\mathcal{H}(\phi(z),\pi(z))}{\delta\pi_{A}(x)\delta\pi_{B}(y)})$
will be added to the action. We can derive this term, through the
action and its functionality of the fields: \be
\frac{\delta^2}{\delta\pi_{A}(x)\delta\pi_{B}(y)}\int
dz\mathcal{H}(\phi,\pi)\mid_{\pi(\phi,\dot{\phi})}=\frac{\delta}{\delta\pi_{A}(x)}\dot{\phi}^B(\phi(y),\pi(y))\mid_{\pi(\phi,\dot{\phi})}
\ee Take attention to $\dot{\phi}^B(x)$ in the right hand side that
is defined independent of $\phi^A(x)$ and is not time derivative.
And also we have the definition of the independent canonical
momentum: \be \pi_{A}(\phi,\dot{\phi})=\frac{\delta
S(\phi,\dot{\phi})}{\delta\dot{\phi}^A} \ee As regards,
$\dot{\phi}^A(x)$ and $\pi_{A}(x)$ are supposed to be independent of
the scalar fields, therefore by use of the chain rule of the
functional derivatives we can write: \be \int dz
\frac{\delta\dot{\phi}^B(\phi(x),\pi(x))}{\delta\pi_{A}(z)}\mid_{\pi(\phi,\dot{\phi})}\times
\frac{\delta\pi_{A}(\phi(z),\dot{\pi}(z))}{\delta\dot{\phi}^C(y)}=\delta^B_{C}\delta(x-y)
\ee and therefore: \be Ln det(\frac{\delta^2\int
dz\mathcal{H}(\phi,\pi)}{\delta\pi_{A}(x)\delta\pi_{B}(y)})=-Ln
det(\frac{\delta^2
S(\phi,\dot{\phi})}{\delta\dot{\phi}^A(x)\delta\dot{\phi}^B(y)}) \ee
If in the left hand side of the present equation, $\pi_{A}(x)$ goes
to the extremum point which is defined in the equation (3.2), thus
$\dot{\phi}_{A}(x)$ in the right hand side presents as the time
derivative of the scalar field. Then by defining the effective
action: \be \Gamma(\varphi(x))=F(J(x))-\int dx J(x)\varphi(x) \ee
and using the equations (3.7) and (3.3), we will find the effective
action at one-loop order: \be
\Gamma^{(1)}(\varphi)=S(\varphi)+\frac{i}{2}tr
Ln(\frac{\delta^2S(\phi)}{\delta\phi^2})\mid_{\varphi}-\frac{i}{2}tr
Ln(\frac{\delta^2S(\phi,\dot{\phi})}{\delta\dot{\phi}^2})\mid_{\dot{\varphi}}
+\delta S^{(1)}(\varphi) \ee The first trace, is defined in the
fourier space as already reminded. And $\delta S^{(1)}(\varphi)$ is
the one-loop ordered counterterm added to the renormalized action.
\section{Renormalization conditions}
Usually, for calculation of the effective action, people have used the assumption that the effective fields are constant. And as a result the continuum matrix inside the trace has been diagonalized. While in our paper the scalar fields cannot be constant values and the derivation of the effective action seems to be difficult.\\
By the way, as an important property, the effective action is a generating function. And we can derive the n-point functions through the functional derivatives of the effective action. An important problem that we encounter in calculating the effective action and n-point functions is renormalization, particularly in the lagrangians which are functions of $H^{AB}$ and have the nonrenormalizable monomials.\\

The first renormalization condition could be assumed is accepting
the common equation of motion between the effective action and the
renormalized action: \be
\frac{\delta\Gamma(\varphi(x))}{\delta\varphi(x)}\mid_{\partial\varphi=b}=0
\ee Inserting (3.9) in (4.1), the present equation will be produced:
\bea
\frac{i}{2}tr[(\frac{\delta^2S}{\delta\varphi^2})^{-1}\frac{\delta}{\delta\varphi(x)}(\frac{\delta^2S}{\delta\varphi^2})]\mid_{\partial\varphi=b}\nonumber\\
-\frac{i}{2}tr[(\frac{\delta^2S}{\delta\dot{\varphi}^2})^{-1}\frac{\delta}{\delta\varphi(x)}(\frac{\delta^2S}{\delta\dot{\varphi}^2})]\mid_{\partial\varphi=b}
+\frac{\delta}{\delta\varphi(x)}\delta
S^{(1)}\mid_{\partial\varphi=b}=0 \eea Since the renormalized action
is a local function, standing continuum matrices like
$G^{AB}_{xy}=(\frac{\delta^2S}{\delta\varphi^A(x)\delta\varphi^B(y)})^{-1}$
on the point $\partial\varphi=b$, leads them to be diagonalized in
the fourier space. for example: \be \alpha(b)p^2G(p)=1 \ee And the
term $\frac{\delta}{\delta\varphi}\delta
S^{(1)}\mid_{\partial\varphi=b}$, simply will be calculated in the
fourier space. And it seems that for the actions with even function
of fields, this term is trivially zero,
$\beta(b)\delta^{\mu}_{A}\int dp\,p_{\mu}=0$. But there are many
different monomials that are not considered in the condition (4.1),
and to determine them we need more conditions like: \be \int
dx_{1}\cdots
dx_{n-1}\frac{\delta^n\Gamma(\varphi)}{\delta\varphi(x_{n})\cdots\delta\varphi(x_{1})}\mid_{\partial\varphi=b}=
\int dx_{1}\cdots dx_{n-1}\frac{\delta^n
S(\varphi)}{\delta\varphi(x_{n})\cdots\delta\varphi(x_{1})}\mid_{\partial\varphi=b}
\ee
And the written equations are trivial for odd n. These equations can be solved in the fourier space because the continuum matrices on the point $\partial\varphi=b$ are diagonal as before.\\

And the point $x_n$ has remaind outside integrals because of the
translational symmetry of the connected and 1PI n-point functions.
It means the same as the momentum conservation in the fourier space.\\

In fact these conditions are written for the 1PI n-point functions
with zero momentum legs in the fourier space. Therefore these
conditions determine the Feynman vertices between interacting
particles which has been
probed by an observer in the IR scale.  \\

If the bare action has the linear symmetry
$\phi^A(x)\longrightarrow\phi^A(x)+a^A$, it is trivial that the
effective action has the same symmetry too. And also we can find the
effective action as a functional of $\partial_{\mu}\varphi^A$, and
the observed object is $\partial_{\mu}\varphi^A$ which is coupled to
the metric.
\subsection{Broken gauge symmetry action for the new degrees of freedom}
Assume a bare action that the renormalization conditions cause the
counterterm monomials which are constructed by both $\varphi^A$ and
$\partial_{\mu}\varphi^A$. In this case, in the points other than
the classical solution, the effective action loses the symmetry
$\varphi(x)\rightarrow\varphi(x)+a$. And even if we have
$\partial_{\mu}\varphi^A$ as classical measurable values, but the
quantum degrees of
freedom of the effective action are values of the scalar fields in all points.\\

And now if we want to have the quantum degrees of freedom analogous
to the classical measurable values in the IR scale, it has been
required that the effective action to be functional of only
$\partial_{\mu}\varphi$ and also to have the symmetry
$\varphi\rightarrow\varphi+a$. For this purpose, the renormalization
conditions have to be redefined. By adding an integral in the free
coordinate $x_{n}$ to the equations (4.4) and choosing a $\varphi$
other than the classical solution, we will have these conditions:
\be \int dx_{1}\cdots
dx_{n}\frac{\delta^n\Gamma(\varphi)}{\delta\varphi(x_{n})\cdots\delta\varphi(x_{1})}=
\int dx_{1}\cdots dx_{n}\frac{\delta^n
S(\varphi)}{\delta\varphi(x_{n})\cdots\delta\varphi(x_{1})}=0 \ee In
which are derived from the Taylor series of $\Gamma(\varphi+a)$
around $a=0$. Indeed we can use the equation $\int
dx\,\frac{\delta\Gamma(\varphi)}{\delta\varphi(x)}=0$ which applied
for the all possible $\varphi(x)$, instead of using the
equations(4.5) that are used for a selected $\varphi(x)$. On this
approach, the symmetry $\varphi\rightarrow\varphi+a$ will be appear
in the effective action again. Now we can write $\Gamma(\varphi)$ as
a functional of $v^A_{\mu}(x)=\partial_{\mu}\varphi^A(x)$ and define
$v^A_{\mu}(x)$ as the effective degrees of freedom. By this
definition, we can have:\be
\frac{\delta^n\Gamma(\varphi)}{\delta\varphi(x_{1})\cdots\delta\varphi(x_{n})}=\partial_{1\mu}\cdots\partial_{n\nu}\frac{\delta^n\Gamma(v)}{\delta
v_{\mu}(x_{1})\cdots\delta v_{\nu}(x_{n})} \ee In which are
consistent to the equations (4.5). It has to be attended to the
renormalization conditions (4.5) that eliminate only the
nonsymmetric monomials. To renormalize the remain monomials, we need
more conditions. Assume that an observer probes the effective vector
fields $v^A_{\mu}(x)$ as the experimental objects. By considering
$v^A_{\mu}(x)=0$ as the expectation values in the effective vacuum
state, we can define the new action for such vector fields. At first
we define the mass $m$ for such vector fields(without gauge
symmetry): \be \int dx e^{i p (x-y)}\frac{\delta^2\Gamma(v)}{\delta
v_{\mu}(x)\delta v_{\nu}(y)}\mid_{p^2=m^2}=0\ee At second, the
coupling constants $g_{n}$ in the new action (interaction amplitudes
between recent experimental objects) will be obtained through the
following equations: \be g_{n}= \int dx_{1}\cdots
dx_{n-1}e^{ip_{1}x_{1}}\cdot
e^{ip_{n-1}x_{n-1}}e^{ip_{n}x_{n}}\frac{\delta^n\Gamma(v)}{\delta
v_{\mu}(x_{1})\cdots\delta
v_{\nu}(x_{n})}\mid_{on-shell}\,\,\,\,n\geq3 \ee Therefore we have
the scalar fields that behave like the massive vector fields in the
IR scale.

If the degrees of freedom like
$g_{\mu\nu}:\eta_{AB}\partial_{\mu}\varphi^A\partial_{\nu}\varphi^B$(one of the metric
$g^{matter}_{\mu\nu}$ introduced by t Hooft) can be defined in a
more advanced way, we can have an arbitrary massive gravity action.
Through choosing the correct mass terms, we can conserve the
unitarity. In fact the one particle state of such an effective
gravity theory will be occurred in the bounding states of the
defined vector field theory.

\section{Conclusion}
The first purpose in this letter, was studding the quantum theory of
the symmetry breaking in the t'Hooft method\cite{Gerard}. In the
path integrals, we separated the usual fourier modes from the
classical solution. By solving the new form of the path integral of
the canonical momentum, we obtained the effective action. Then the
renormalization conditions were implemented on the effective
action.\\

The next purpose, was finding the new degrees of freedom
corresponded to the coordinate derivative of the fields. Such new
degrees of freedom define a quantum massive vector field theory in
the IR scale. Through
studding the new renormalization conditions, we discussed on the recent purpose.\\

From now on, one can select a real sample in the actions to continue
more practically these present computations. And also, the effective
action of the gravity could be calculated and added to the present
effective action of the scalar fields. As well, one can obtain the
new degrees of freedom from the original scalar fields to offers
them as the quantum massive gravity in the IR scale.

\vspace*{1cm}

{\bf Acknowledgments}

I would like to thank dear Navid Abbasi for his useful discussions
and comments.

\end{document}